\documentclass[runningheads]{llncs}
\usepackage{cite}
\usepackage{graphicx}
\usepackage{hyperref}
\usepackage{url}

\usepackage{amsmath}
\usepackage{amssymb}

\usepackage{multirow}
\usepackage{array}
\newcolumntype{P}[1]{>{\raggedright\arraybackslash}p{#1}}
\usepackage{tabu}

\usepackage[labelfont=bf]{caption}
\usepackage{subcaption}

\usepackage{colortbl} 
\usepackage[table]{xcolor}
\usepackage{tikz}
\usetikzlibrary{arrows,backgrounds,calc,decorations.pathmorphing,decorations.pathreplacing,fit,matrix,patterns,petri, positioning, shapes,shapes.multipart,matrix,positioning,shapes.callouts,shapes.arrows,calc, decorations.shapes}
\tikzset{
	silent/.style={
		minimum width{=}2mm,
		minimum height{=}5mm,
		fill{=}gray
	}
}

\usepackage[ruled,lined,vlined]{algorithm2e}

\SetCommentSty{xCommentSty}

\usepackage{multicol}

\newcommand{\activity}{\ensuremath{a}}
\newcommand{\alignment}{\ensuremath{\gamma}}
\newcommand{\caseId}{\ensuremath{c}}

\newcommand{\dataStructure}{\ensuremath{\mathcal{D}}}
\newcommand{\event}{\ensuremath{e}}

\newcommand{\labelFunc}{\ensuremath{\lambda}}
\newcommand{\marking}{\ensuremath{M}}
\newcommand{\mset}{\ensuremath{B}}
\newcommand{\naturals}{\ensuremath{\mathbb{N}}}
\newcommand{\naturalsZero}{\ensuremath{\naturals_0}}
\newcommand{\petriNet}{\ensuremath{N}}
\newcommand{\place}{\ensuremath{p}}
\newcommand{\places}{\ensuremath{P}}
\newcommand{\pnArcs}{\ensuremath{F}}
\newcommand{\prefixAlignment}{\ensuremath{\overline{\alignment}}}

\newcommand{\proj}{\ensuremath{\pi}}

\newcommand{\reachableMarkings}{\ensuremath{\mathcal{R}}}
\newcommand{\reals}{\ensuremath{\mathbb{R}}}
\newcommand{\sequence}{\ensuremath{\sigma}}

\newcommand{\skipAct}{\gg}
\newcommand{\stream}{\ensuremath{S}}
\newcommand{\synchronousProduct}{\ensuremath{N^S}}

\newcommand{\transition}{\ensuremath{t}}
\newcommand{\transitions}{\ensuremath{T}}
\newcommand{\univAct}{\ensuremath{\mathcal{A}}}

\newcommand{\univCase}{\ensuremath{\mathcal{C}}}
\newcommand{\univEvent}{\ensuremath{\mathcal{E}}}
\newcommand{\univMSet}{\ensuremath{{\mathcal{B}}}}

\definecolor{logMove}{HTML}{333333}
\definecolor{synchronousMove}{HTML}{999999}

\newcolumntype{l}{>{\columncolor{logMove}\color{white}}c}
\newcolumntype{s}{>{\columncolor{synchronousMove}}c}

\newcolumntype{?}{!{\vrule width 1.5pt}}

\begin{document}
\title{Online Process Monitoring Using Incremental State-Space Expansion: An Exact Algorithm}
%
\titlerunning{Online Process Monitoring Using Incremental State-Space Expansion}
%
\author{Daniel Schuster\inst{1} \and
Sebastiaan J. van Zelst\inst{1,2}}
\authorrunning{D. Schuster, S. J. v. Zelst}
%
\institute{ Fraunhofer Institute for Applied Information Technology FIT, \\Sankt Augustin, Germany\\ 
\email{\{daniel.schuster,sebastiaan.van.zelst\}@fit.fraunhofer.de}\\
\and
RWTH Aachen University, Aachen, Germany\\
\email{s.j.v.zelst@pads.rwth-aachen.de}
}
\maketitle              

\begin{abstract}

The execution of (business) processes generates valuable tra-ces of event data in the information systems employed within companies. 
Recently, approaches for monitoring the correctness of the execution of running processes have been developed in the area of process mining, i.e., online conformance checking. 
The advantages of monitoring a process' conformity during its execution are clear, i.e., deviations are detected as soon as they occur and countermeasures can immediately be initiated to reduce the possible negative effects caused by process deviations. 
Existing work in online conformance checking only allows for obtaining approximations of non-conformity, e.g., overestimating the actual severity of the deviation.
In this paper, we present an exact, parameter-free, online conformance checking algorithm that computes conformance checking results on the fly. 
Our algorithm exploits the fact that the conformance checking problem can be reduced to a shortest path problem, by incrementally expanding the search space and reusing previously computed intermediate results. 
Our experiments show that our algorithm is able to outperform comparable state-of-the-art approximation algorithms.

\end{abstract}
\section{Introduction}
Modern information systems support the execution of different business processes within companies.
Valuable traces of \emph{event data}, describing the various steps performed during process execution, are easily extracted from such systems.
The field of \emph{process mining}~\cite{DBLP:books/sp/Aalst16} aims to exploit such information, i.e., the event data, to better understand the overall execution of the process.
For example, in process mining, several techniques have been developed that allow us to \emph{(i)} automatically discover process models, \emph{(ii)} compute whether the process, as reflected by the data, conforms to a predefined reference model and \emph{(iii)} detect performance deficiencies, e.g., bottleneck detection.

\begin{figure}[tb]
    \centering
    \includegraphics[width=\textwidth, trim={0cm 10.4cm 4.1cm 0cm},clip]{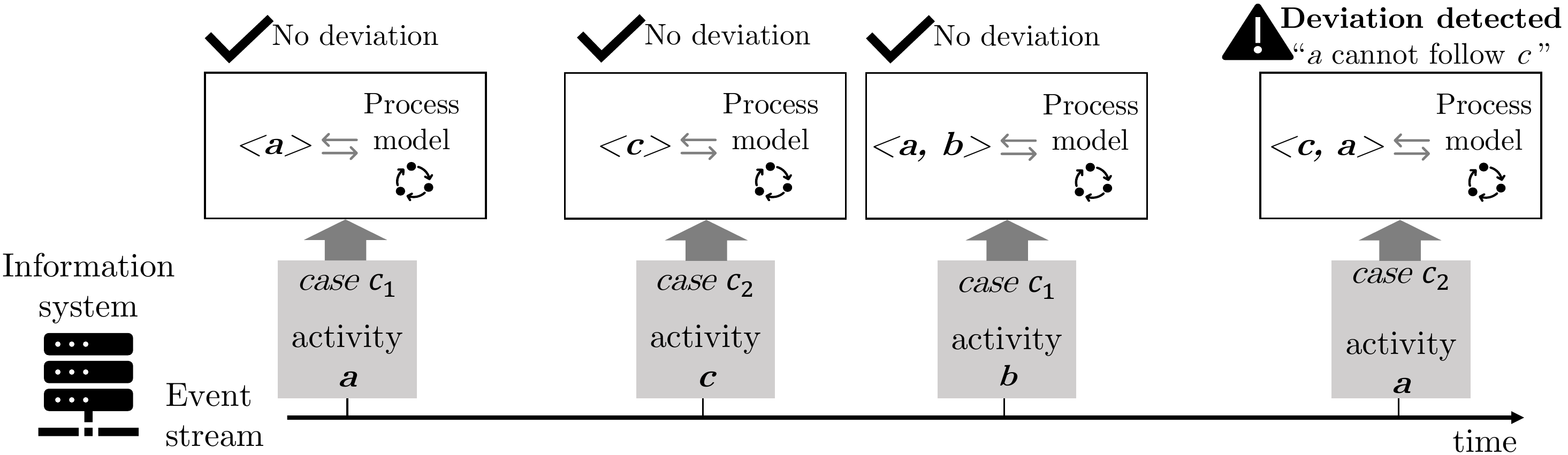}
    \caption{Overview of online process monitoring. Activities are performed for different process instances, identified by a case-id, over time. Whenever a new activity is executed, the sequence of already executed activities within the given case/process instance is checked for conformance w.r.t. a reference process model}
    \label{fig:overview_online_conformance}
\end{figure}
 
The majority of existing process mining approaches work in an offline setting, i.e., data is captured over time  during process execution and process mining analyses are performed a posteriori.
However, some of these techniques benefit from an \emph{online application scenario}, i.e., analyzing the process at the moment it is executed.
Reconsider \emph{conformance checking}, i.e., computing whether a process' execution conforms to a reference model.
Checking conformance in an online setting allows the process owner to detect and counteract non-conformity at the moment it occurs (Fig.~\ref{fig:overview_online_conformance}).
Thus, potential negative effects caused by a process deviation can be mitigated or eliminated.
This observation inspired the development of novel conformance checking algorithms working in an online setting~\cite{vanZelst2017_ijdsa,DBLP:conf/bpm/Burattin17,DBLP:conf/bpm/BurattinZADC18}.
However, such algorithms provide approximations of non-conformity and/or use high-level abstractions of the reference model and the event data, i.e., not allowing us to obtain an exact quantification of non-conformance.

In this paper, we propose a novel, \emph{exact} solution for the online conformance checking problem.
We present a \emph{parameter-free} algorithm that computes exact conformance checking results and provides an exact quantification of non-conformance.
Our algorithm exploits the fact that the computation of conformance checking results can be reduced to a shortest path problem.
In fact, we extend the search space in the course of a process instance execution and compute shortest paths by utilizing previous results every time new behavior is observed. 
Moreover, we explicitly exploit specific properties of the search space when solving the conformance checking problem.
Therefore, the proposed incremental algorithm is specifically designed for online conformance checking and cannot be directly applied to general shortest path problems.
The conducted experiments show that the proposed approach outperforms existing approximation algorithms and additionally guarantees exact results.

The remainder of this paper is structured as follows.
In Section~\ref{sec:related_work}, we present related work regarding conformance checking and incremental search algorithms. 
In Section~\ref{sec:prelim}, we present preliminaries. 
In Section~\ref{sec:algo}, we present the main algorithm.
In Section~\ref{sec:correctness}, we prove the correctness of the proposed algorithm.
We evaluate the proposed algorithm and present the results of the experiments conducted in Section \ref{sec:eval}.
Finally, we conclude the paper in Section~\ref{sec:conclusion}.

\section{Related Work}
\label{sec:related_work}
In this section, we first focus on (online) conformance checking techniques. 
Subsequently, we present related work regarding incremental search algorithms.

Two early techniques, designed to compute conformance statistics, are token-based replay~\cite{DBLP:journals/is/RozinatA08} that tries to replay the observed behavior on a reference model and footprint-based comparison~\cite{DBLP:books/sp/Aalst16}, in which the event data and the process model are translated into the same abstraction and then compared.
As an alternative, \emph{alignments} have been introduced~\cite{DBLP:journals/widm/AalstAD12,adriansyah_2014_phd_aligning} that map the observed behavioral sequences to a feasible execution sequence as described by the (reference) process model. 
Alignments indicate whether behavior is missing and/or whether inappropriate behavior is observed. 
The problem of finding an alignment was shown to be reducible to the shortest path problem~\cite{adriansyah_2014_phd_aligning}. 

The aforementioned techniques are designed for offline usage, i.e., they work on static (historical) data.
In \cite{vanZelst2017_ijdsa}, an approach is presented to monitor ongoing process executions based on an event stream by computing partially completed alignments each time a new event is observed.
The approach results in approximate results, i.e., false negatives occur w.r.t. deviation detection. 
In this paper, we propose an approach that extends and improves~\cite{vanZelst2017_ijdsa}.
In~\cite{DBLP:conf/bpm/Burattin17}, the authors propose to pre-calculate a transition system that supports replay of the ongoing process. 
Costs are assigned to the transition system's edges and replaying a deviating process instance leads to larger (non-zero) costs.
Finally, \cite{DBLP:conf/bpm/BurattinZADC18} proposes to compute conformance of a process execution based on all possible behavioral patterns of the activities of a process. 
However, the use of such patterns leads to a loss of expressiveness in deviation explanation and localization.


In general, incremental search algorithms find shortest paths for similar search problems by utilizing results from previously executed searches~\cite{DBLP:journals/aim/KoenigLLF04}. 
In~\cite{DBLP:conf/nips/KoenigL01}, the Lifelong Planning $A^*$ algorithm is introduced that is an incremental version of the $A^*$ algorithm. 
The introduced algorithm repeatedly calculates a shortest path from a fixed start state to a fixed goal state while the edge costs may change over time.
In contrast, in our approach, the goal states are constantly changing in each incremental execution, whereas the edge costs remain fixed.
Moreover, only new edges and vertices are incrementally added, i.e., the already existing state space is only extended.
In~\cite{DBLP:conf/atal/KoenigL05}, the Adaptive $A^*$ algorithm is introduced, which is also an incremental version of the $A^*$ algorithm.
The Adaptive $A^*$ algorithm is designed to calculate a shortest path on a given state space from an incrementally changing start state to a fixed set of goal states.
In contrast to our approach, the start state is fixed in each incremental execution.

\section{Background}
\label{sec:prelim}
In this section, we present basic notations and concepts used within this paper.

Given a set $X$, a multiset $\mset$ over $X$ allows us to assign a multiplicity to the elements of $X$, i.e., $\mset \colon X {\to} \naturalsZero$.
Given $X {=}\{x,y,z\}$, the multiset $[x^5,y]$ contains $5$ times $x$, once $y$ and no $z$. 
The set of all possible multisets over a set $X$ is denoted by $\univMSet(X)$.
We write $x {\in_+} B$ if $x$ is contained at least once in multiset $B$.

A sequence $\sequence$ of length $n$, denoted by $|\sequence|{=}n$, over a base set $X$ assigns an element to each index, i.e., $\sequence {\colon} \{1,\dots,n\} {\to} X$.
We write a sequence $\sequence$ as $\langle \sequence(1), \sequence(2), ..., \sequence(|\sequence|)\rangle$.
Concatenation of sequences is written as $\sequence {\cdot} \sequence'$, e.g., $\langle x,y\rangle {\cdot}\allowbreak \langle z \rangle{=}\allowbreak\langle x,y,z \rangle$.
The set of all possible sequences over base set $X$ is denoted by $X^*$.
For element inclusion, we overload the notation for sequences, i.e., given $\sequence{\in}X^*$ and $x{\in}X$, we write $x{\in}\sequence$ if $\exists \: 1 {\leq} i {\leq} |\sequence| \left(\sequence(i){=}x\right)$, e.g., $b{\in}\langle a,b\rangle$. 

Let $\sequence{\in}X^*$ and let $X'{\subseteq}{X}$.
We recursively define $\sequence_{\downarrow_{X'}}{\in}X'^*$ with:
$\langle\rangle_{\downarrow_{X'}} {=} \langle\rangle$, $(\langle x\rangle {\cdot} \sequence)_{\downarrow_{X'}}{=} \langle x \rangle {\cdot} \sequence_{\downarrow_{X'}}$ if $x{\in}X'$ and $(\langle x\rangle {\cdot} \sequence)_{\downarrow_{X'}}{=} \sequence_{\downarrow_{X'}}$ if $x{\notin}X'$.
For example, let $X'{=}\{a,b\}$, $X{=}\{a,b,c\}$, $\sequence {=} \langle a,c,b,a,c \rangle {\in} X^*$ then $\sequence_{\downarrow_{X'}}{=} \langle a,b,a \rangle$.

Let $t{=}(x_1,...,x_n){\in}X_1 {\times} {\cdots} {\times} X_n$ be an $n$-tuple, we let $\proj_1(t){=}x_1,\allowbreak \dots,\allowbreak\proj_n(t){=}x_n$ denote the corresponding projection functions that extract a specific component from the tuple, e.g., $\proj_3((a,b,c)){=}c$.
Correspondingly, given a sequence $\sequence{=}\langle(x^1_1,\dots,x^1_n),\dots,(x_1^m,\dots,x_n^m)\rangle$ with length $m$ containing $n$-tuples, we define projection functions $\proj^*_1(\sequence){=}\langle x_1^1,\allowbreak\dots,x_1^m\rangle,\allowbreak\dots,\allowbreak\proj^*_n(\sequence){=}\langle x_n^1,\dots,x_n^m\rangle$ that extract a specific component from each tuple and concatenate it into a sequence.
For instance, $\proj^*_2 \left( \langle  \left(a,b\right), \left(c,d\right), \left(c,b\right) \rangle \right) {=} \langle b,d,b \rangle$.

\subsubsection{Event Logs}

\begin{figure}[tb]
\begin{minipage}{\textwidth}
  \begin{minipage}[]{0.55\textwidth}
    \scriptsize
    \captionof{table}{Example \emph{event log} fragment}
    \label{tab:event_log}
        \begin{tabular}{|c|c|c|c|}
			\hline
			\textbf{Case} & \textbf{Activity} & \textbf{Resource} & \textbf{Time-stamp} \\
			\hline
			$\cdots$ & $\cdots$ & $\cdots$ & $\cdots$ \\
			\textit{13152} & create account $(a)$ & \textit{Wil} & \textit{19-04-08 10:45} \\
			\textit{13153} & create account $(a)$ & \textit{Bas} & \textit{19-04-08 11:12} \\
			\textit{13154} & request quote $(c)$ & \textit{Daniel} & \textit{19-04-08 11:14} \\
			\textit{13155} & request quote $(c)$ & \textit{Daniel} & \textit{19-04-08 11:40} \\
			\textit{13152} & submit order $(b)$ & \textit{Wil} & \textit{19-04-08 11:49} \\
			$\cdots$ & $\cdots$ & $\cdots$ & $\cdots$ \\
			\hline
		\end{tabular}
  \end{minipage}
  \hfill
  \begin{minipage}[]{0.38\textwidth}
    \scriptsize
    \begin{tikzpicture}
        \draw[line width=1pt, |->, >=latex'](0,0) -- coordinate (x axis) (4.7,0) node[below left] {time}; 
        \node[] at (2.35,0.16) {$(13152,a),(13153,a),(13154,c),\cdots$};
    \end{tikzpicture}
      \captionof{figure}{Schematic example of an event stream}
      \label{fig:example_stream}
    \end{minipage}
  \end{minipage}
 \end{figure}

The data used in process mining are \emph{event logs}, e.g., consider Tab.~\ref{tab:event_log}. 
Each row corresponds to an \emph{event} describing the execution of an activity in the context of an \emph{instance} of the process.
For simplicity, we use short-hand activity names, e.g., $a$ for ``create account''.
The events related to \emph{Case-id} $13152$ describe the activity sequence $\langle a,b \rangle$.

\subsubsection{Event streams}
In this paper, we assume an \emph{event stream} rather than an event log. 
Conceptually, an event stream is an (infinite) sequence of events.
In Fig.~\ref{fig:example_stream}, we depict an example. 
For instance, the first event, $(13152,a)$, indicates that for a process instance with case-id $13152$ activity $a$ was performed. 

\begin{definition}[Event; Event Stream]
\label{def:event_stream}
Let $\univCase$ denote the universe of \emph{case identifiers} and $\univAct$ the universe of \emph{activities}. 
An event $\event{\in}\univCase{\times}\univAct$ describes the execution of an activity $\activity{\in}\univAct$ in the context of a process instance identified by $\caseId{\in}\univCase$.
An event stream $\stream$ is a sequence of events, i.e., $\stream{\in}(\univCase{\times}\univAct)^*$.
\end{definition}
As indicated in Table~\ref{tab:event_log}, real-life events contain additional information, e.g., resource information, and are usually uniquely identifiable by an event id. 
However, for the purpose of this paper, we are only interested in the executed activity, the case-id of the corresponding process instance and the order of events.

\subsubsection{Process Models}
\begin{figure}[tb]
	\centering
        \scriptsize
    	\begin{tikzpicture}[node distance=1.3cm,>=stealth',bend angle=20,auto]
     		\tikzstyle{place}=[circle,thick,draw=black,minimum size=5mm]
    	  	\tikzstyle{transition}=[thick,draw=black,minimum size=5mm]
    	  	\tikzstyle{silent}=[rectangle,thick,draw=black,fill=black,minimum size=5mm, text=white]
    	    \node [place,tokens=1,label=below:{$p_1$}] (p1) {};
    		\node [transition] (A) [right of = p1, label=below:{$t_1$}, label={[align=center]above:create account}]  {$a$};    
    	    \node [place] (p2) [right of = A,label=below:{$p_2$}] {};
    		\node [silent] (tau) [below of =A, label=above:{$t_2$}]  {$\tau$};
    		\node [transition] (B) [right of= p2,label=below:{$t_3$},label={[align=center]above:submit order}]  {$b$};
    		\node [transition] (C) [below of= B,label=above:{$t_4$},label={[align=center]below:request quote}]  {$c$};
    	    \node [place] (p3) [right of= B,label=below:{$p_3$}] {};
    		\draw [->] (p1) to (A); 
    		\draw [->] (A) to (p2); 
    		\draw [->] (p2) to (B); 
    		\draw [->] (p2) to (C); 
    		\draw [->] (B) to (p3); 
    		\draw [->] (C) to (p3); 
    		
    		\draw [->] (p1) to (tau);
    		\draw [->] (tau) to (p2);
	    \end{tikzpicture}
	\caption{Example WF-net $\petriNet_1$ with visualized initial marking $[\place_1]$ and final marking $[\place_3]$ describing a simplified ordering process. First ``create account'' is optionally executed. Next, either ``submit order'' or ``request quote'' is executed}
	\label{fig:example_petri_net}
\end{figure}
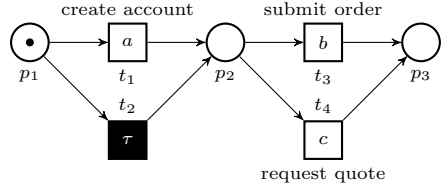

Process models allow us to describe the (intended) behavior of a process.
In this paper, we focus on \emph{sound Workflow nets}~\cite{DBLP:journals/jcsc/Aalst98}.
A Workflow net (WF-net) is a subclass of \emph{Petri nets}~\cite{murata_1989}.
Sound WF-nets, in turn, are a subclass of WF-nets with favorable \emph{behavioral properties}, e.g., no deadlocks and live-locks.
Consider Fig.~\ref{fig:example_petri_net}, where we depict a sound WF-net.
We use WF-nets since many high-level process modeling formalism used in practice, e.g. BPMN~\cite{DBLP:journals/infsof/DijkmanDO08}, are easily translated into WF-nets.
Moreover, it is reasonable to assume that an experienced business process designer creates sound process models. 

Petri nets consist of a set of \emph{places} $\places$, visualized as circles, and a set of \emph{transitions} $\transitions$, visualized as rectangular boxes.
Places and transitions are connected by arcs which are defined by the set $\pnArcs{=}(\places{\times}\transitions){\cup}(\transitions{\times}\places)$.
Given an element $x{\in}\places{\cup}\transitions$, we write $x{\bullet} {=} \left\{y{\in}\places{\cup}\transitions \mid (x,y){\in}\pnArcs\right\}$ to define all elements $y$ that have an incoming arc from $x$.
Symmetrically, ${\bullet} x{=}\left\{y{\in}\places{\cup}\transitions \mid (y,x){\in}\pnArcs\right\}$, e.g., ${\bullet} p_2{=}\left\{t_1,t_2\right\}$ (Fig. \ref{fig:example_petri_net}).

The state of a Petri net, i.e., a \emph{marking} $\marking$, is defined by a multiset of places, i.e., $\marking {\in} \univMSet(\places)$.
Given a Petri net $\petriNet$ with a set of places $\places$ and a marking $\marking{\in}\univMSet(\places)$, a \emph{marked net} is written as $(\petriNet,\marking)$.
We denote the initial marking of a Petri net with $\marking_i$ and the final marking with $\marking_f$.
We denote a Petri net as $\petriNet{=}(\places,\transitions,\pnArcs,\marking_i,\marking_f, \labelFunc)$. 
The labeling function $\lambda{\colon}\transitions{\to}\univAct{\cup}\{\tau\}$ assigns an (possibly invisible, i.e., $\tau$) activity label to each transition, e.g., $\lambda(\transition_1){=}a$ in Fig.~\ref{fig:example_petri_net}.

The transitions of a Petri net allow to change the state.
Given a marking $\marking{\in}\univMSet(\places)$, a transition $\transition$ is \emph{enabled} if $\forall \place{\in}{\bullet}\transition\left(\marking(\place){>}0\right)$.
An enabled transition can \emph{fire}, yielding marking $\marking'{\in}\univMSet(\places)$, where $\marking'(\place){=}\marking(\place){+}1$ if $\place{\in}\transition {\bullet} {\setminus} {\bullet} \transition$, $\marking'(\place){=}\marking(\place){-}1$ if $\place{\in}{\bullet} \transition {\setminus} \transition {\bullet}$, otherwise $\marking'(\place){=}\marking(\place)$.
We write $(\petriNet,\marking)[\transition\rangle$ if $\transition$ is enabled in $\marking$ and we write $(\petriNet,\marking){\xrightarrow{\transition}}(\petriNet,\marking')$ to denote that firing transition $\transition$ in marking $\marking$ yields marking $\marking'$.
In Fig.~\ref{fig:example_petri_net}, we have $(\petriNet_1,[\place_1])[\transition_1\rangle$ as well as $(\petriNet_1,[\place_1]){\xrightarrow{\transition_1}}(\petriNet_1,[\place_2])$.
If a sequence of transitions $\sequence{\in}\transitions^*$ leads from marking $\marking$ to $\marking'$, we write $(\petriNet,\marking) {\xrightarrow{\sequence}} (\petriNet,\marking')$.
We let $\reachableMarkings(\petriNet,\marking){=}\{\marking'{\in}\univMSet(\places) \mid \exists\sequence{\in}\transitions^* (\petriNet,\marking) {\xrightarrow{\sequence}} (\petriNet,\marking') \}$ denote the state space/all reachable markings of $N$ given an initial marking $M$.

A \emph{WF-net} $N{=}(\places,\transitions,\pnArcs,[\place_i],[\place_o],\labelFunc)$ is a Petri net with a unique source place $p_i$ and a unique sink place $p_o$, i.e., $\marking_i{=}[p_i]$ and $\marking_f{=}[p_o]$.
Moreover, every element $x{\in}\places{\cup}\transitions$ is on a path from $\place_i$ to $\place_o$.

\subsubsection{Alignments} 

\begin{figure}[tb]
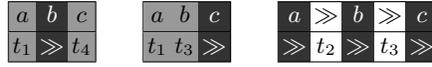

    \footnotesize
    \centering
        {
        \begin{tabular}{|s l s|}\hline
        $a$     & $b$     & $c$     \\ \hline
        $t_1$   & $\gg$     & $t_4$   \\ \hline
        \end{tabular}
        \hspace{0.4cm}
        \begin{tabular}{|s s l|}\hline
        $a$     & $b$     & $c$     \\ \hline
        $t_1$   & $t_3$     & $\gg$   \\ \hline
        \end{tabular}
        \hspace{0.4cm}
        \begin{tabular}{|l c l c l|}\hline
        $a$     & $\gg$     & $b$   & $\gg$ & $c$   \\ \hline
        $\gg$   & $t_2$     & $\gg$ & $t_3$ & $\gg$  \\ \hline
        \end{tabular}
        }
    \caption{Three possible alignments for WF-net $N_1$ (Fig.~\ref{fig:example_petri_net}) and trace $\langle a,b,c \rangle$}
    \label{fig:alignments}
\end{figure}
To explain traces in an event log w.r.t. a reference model, we use \emph{alignments}~\cite{adriansyah_2014_phd_aligning}, which map a trace onto an execution sequence of a model.
Exemplary alignments are depicted in Fig.~\ref{fig:alignments}.
The first row of an alignment (ignoring the skip symbol $\gg$) equals the trace and the second row (ignoring $\gg$) represents a sequence of transitions leading from the initial to the final marking.

We distinguish three types of \emph{moves} in an alignment. 
A \emph{synchronous move} (light-gray) matches an observed activity to the execution of a transition, where the transition's label must match the activity. 
\emph{Log moves} (dark-gray) indicate that an activity is not re-playable in the current state of the process model. 
\emph{Model moves} (white) indicate that the execution of a transition cannot be mapped onto an observed activity.
They can be further differentiated into \emph{invisible}- and \emph{visible model moves}. 
An invisible model move consists of an inherently invisible transition ($\labelFunc(\transition) {=} \tau$). 
Visible model moves indicate that an activity should have taken place w.r.t the model but was not observed at that time.

In an online setting, an event stream is assumed to be infinite. 
A new event for a given process instance can occur at any time.
Hence, we are interested in explanations of the observed behavior that still allow us to reach the final state in the reference model, i.e., \emph{prefix-alignments}. 
The first row of a prefix-alignment also corresponds to the trace, but the second row corresponds to a sequence of transitions leading from the initial marking to a marking from which the final marking can still be reached.
For a formal definition, we refer to~\cite{adriansyah_2014_phd_aligning}.

Since multiple (prefix-)alignments exist, we are interested in an alignment that minimizes the mismatches between the trace and the model.
Therefore, we assign costs to moves.
We use the \emph{standard cost function}, which assigns cost $0$ to synchronous moves and invisible model moves, and cost $1$ to log- and visible model moves.
A \mbox{(prefix-)alignment} is optimal if it has minimal costs.


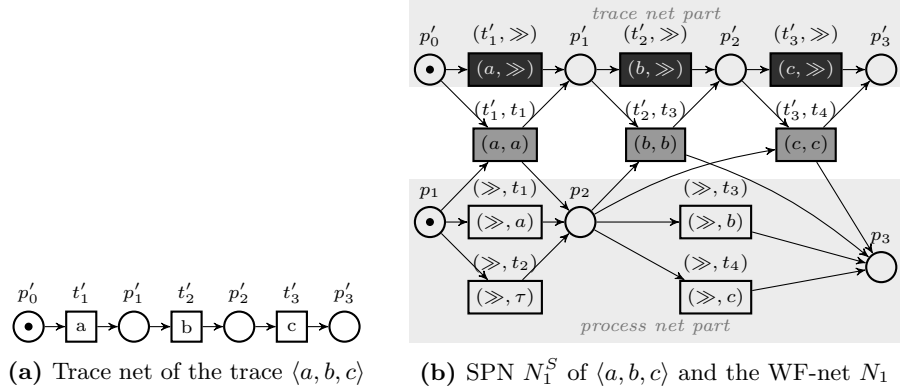
\begin{figure}[tb]
    \centering
    \begin{subfigure}[t]{0.39\textwidth}
        \centering
    	\scriptsize
    	\begin{tikzpicture}[node distance=.7cm and 2cm,>=stealth',auto]
    	  \tikzstyle{place}=[circle,thick,draw=black,minimum size=4mm]
    	  \tikzstyle{transition}=[thick,draw=black,minimum size=4mm]
    	  \tikzstyle{silent}=[rectangle,thick,draw=black,fill=black,minimum size=6mm, text=white]
    	    \node [place,tokens=1,label={[name=l1] $p_0'$}] (p0) {};
    	    \node [transition, label=$t_1'$, right of = p0] (a) {a};
    	    \draw[->] (p0) to (a);
    	    \node[place,label=$p_1'$, right of=a] (p1) {};
    	    \draw[->] (a) to (p1);
    	    \node [transition, label=$t_2'$, right of = p1] (b) {b};
    	    \draw[->] (p1) to (b);
    	    \node[place,label=$p_2'$, right of=b] (p2) {};
    	    \draw[->] (b) to (p2);
    	    \node [transition, label=$t_3'$, right of = p2] (c) {c};
    	    \draw[->] (p2) to (c);
    	    \node[place,label=$p_3'$, right of=c] (p3) {};
    	    \draw[->] (c) to (p3);
    	\end{tikzpicture}
    	\caption{Trace net of the trace $\langle a, b, c \rangle$}
    	\label{fig:trace_net}
    \end{subfigure}
    \hfill
    \begin{subfigure}[t]{0.59\textwidth}
        \centering
    	\scriptsize
    	\begin{tikzpicture}[node distance=1cm and 1.5cm,>=stealth',bend angle=10,auto]
    	  \tikzstyle{place}=[circle,thick,draw=black,minimum size=4mm]
    	  \tikzstyle{transition}=[thick,draw=black,minimum size=4mm]
    	  \tikzstyle{silent}=[rectangle,thick,draw=black,fill=black,minimum size=6mm, text=white]
    	
    	    \node [place,tokens=1,label={[name=l1] $p_0'$}] (p0_) {};
    		\node [transition] (t1_) [right of = p0_, label={[name=l] $(t_1',\gg)$}, fill=logMove]  {$\color{white}(a,\gg)$};
    		\node [place,tokens=0,label={$p_1'$}, right of = t1_] (p1_) {};
    		\node [transition] (t2_) [right of = p1_, label={$(t_2',\gg)$}, fill=logMove]  {\color{white}$(b,\gg)$};  
    		\node [place,tokens=0,label={$p_2'$}, right of = t2_] (p2_) {};
    		\node [transition] (t3_) [right of = p2_, label={$(t_3',\gg)$}, fill=logMove]  {\color{white}$(c,\gg)$};  
    		\node [place,tokens=0,label={[name=3] $p_3'$}, right of = t3_] (p3_) {};
    		\draw [->,] (p0_) to (t1_); 
    		\draw [->,] (t1_) to (p1_); 
    		\draw [->,] (p1_) to (t2_); 
    		\draw [-> ] (t2_) to (p2_); 
    		\draw [->,] (p2_) to (t3_); 
    		\draw [->,] (t3_) to (p3_);
    		\node [transition,label={$(t_1',t_1)$}, below of = t1_, fill=synchronousMove] (t_aa) {$(a,a)$};
    		\draw [->,] (p0_) to (t_aa); 
    		\draw [->,] (t_aa) to (p1_); 
    
    		\node [transition,label={$(t_2',t_3)$}, below of = t2_, fill=synchronousMove] (t_bb) {$(b,b)$};
    		\draw [->,] (p1_) to (t_bb); 
    		\draw [->,] (t_bb) to (p2_); 
    		
    		\node [transition,label={$(t_3',t_4)$}, below of = t3_, fill=synchronousMove] (t_cc) {$(c,c)$};
    		\draw [->,] (p2_) to (t_cc); 
    		\draw [->,] (t_cc) to (p3_); 
    		
    		\node [place,label={[name=l3] $p_1$}, below = 1.6cm of  p0_,tokens=1] (p1) {};
    		\node [transition] (A) [right of = p1, label={$(\gg,t_1)$}]  {$(\gg,a)$};    
    	    \node [place] (p2) [right of = A,label={$p_2$}] {};
    		\node [transition] (B) [right= .5cm and 1.1cm of p2,label={$(\gg,t_3)$}]  {$(\gg,b)$};y
    		
    		\node [transition] (tau) [below of = A,label={$(\gg,t_2)$}]  {$(\gg,\tau)$};
    		
    		\node [transition] (C) [below of = B,label={$(\gg,t_4)$}]  {$(\gg,c)$};
    	    \node [place] (p3) [below = 2.2cm of  p3_,label={$p_3$}] {};
    		\draw [->] (p1) to (A); 
    		\draw [->] (A) to (p2); 
    		\draw [->] (p2) to (B); 
    		\draw [->] (p1) to (tau); 
    		\draw [->] (p2) to (C); 
    		\draw [->] (B) to (p3); 
    		\draw [->] (C) to (p3); 
    		\draw [->] (tau) to (p2); 
    		
    		\draw [->] (p1) to (t_aa);
    		\draw [->] (t_aa) to (p2);
    		
    		\draw [->] (p2) to (t_bb);
    		\draw [->,bend left] (t_bb) to (p3);
    		
    		\draw [->,bend left] (p2) to (t_cc);
    		\draw [->] (t_cc) to (p3);
    		
    		\node[above =.3cm of t2_,color=gray](tracenet){\itshape trace net part};
    		\node[below =3cm of t2_,color=gray](processnet){\textit{process net part}};
    		
    		\node[fill=black, opacity=0.075,fit=(p0_) (p3_) (l1) (tracenet),inner xsep=1pt,inner ysep=0.2pt] {};
    		\node[fill=black, opacity=0.075,fit=(p1) (p3) (C) (l3) (processnet),inner xsep=1pt,inner ysep=0.1pt] {};
    	\end{tikzpicture}
		\caption{SPN $\synchronousProduct_1$ of $\langle a, b, c \rangle$ and the WF-net $\petriNet_1$}
		\label{fig:sync_product_net}
    \end{subfigure}
    \caption{Construction of a trace net and a synchronous product net (SPN)}
\end{figure}

To compute an optimal (prefix-)alignment, we search for a \emph{shortest path} in the state-space of the \emph{synchronous product net (SPN)}~\cite{adriansyah_2014_phd_aligning}.
An SPN is composed of a trace net and the given WF-net. 
In Fig.~\ref{fig:trace_net}, we depict an example trace net.
We refer to~\cite{adriansyah_2014_phd_aligning} for a formal definition of the trace net.
In Fig.~\ref{fig:sync_product_net} we depict an example SPN. 
Each transition in the SPN corresponds to a (prefix-)alignment move.
Hence, we can assign costs to each transition.
Any path in the state-space (sequence of transitions in the SPN) from $[\place'_0, \place_1]$ to $[\place'_3,\place_3]$ corresponds to an alignment of $N_1$ and $\langle a,b,c \rangle$.
For the given example, a shortest path with cost 1, which corresponds to the first alignment depicted in Fig.~\ref{fig:alignments}, is:
$(N_1^S,[\place'_0, \place_1]) {\xrightarrow{(t_1',t_1)}} (N_1^S,[p_1',p_2]) {\xrightarrow{(t_2',\gg)}} (N_1^S,[p_2',p_2]) {\xrightarrow{(t_3',t_4)}} (N_1^S,[p_3',p_3])$
\\To compute a \emph{prefix-alignment}, we look for a shortest path from the initial marking to a marking $\marking{\in} \reachableMarkings(\synchronousProduct_1, [p_0',p_1])$ such that $\marking(p_3'){=}1$, i.e., the last place of the trace net part is marked. 
Next, we formally define the SPN. 

\begin{definition}[Synchronous Product Net (SPN)]
For a given trace $\sequence$, the corresponding trace net $N^{\sequence} {=} (P^{\sequence},T^{\sequence},F^{\sequence},[p_i^{\sequence}],[p_o^{\sequence}],\lambda^{\sequence})$ and a WF-net $N {=} (\places,\transitions,\allowbreak \pnArcs,[\place_i],[\place_o],\labelFunc)$ s.t. $P^\sequence {\cap} P {=}\emptyset$ and $F^\sequence {\cap} F{=}\emptyset$, we define the SPN $N^{S} {=} (P^{S},T^{S},\allowbreak F^{S},\marking_i^{S},\marking_f^{S},\lambda^{S})$ s.t.:
\begin{itemize}

\item $P^S {=} P^{\sequence} {\cup} P$

\item $T^{S}{=}( T^\sequence {\times} \{\gg \}) \cup (\{\gg \}{\times} T) \cup \{ (t',t) {\in}  T^\sequence{\times} T \mid \lambda(t){=}\lambda^\sequence(t'){\neq} \tau \}$

\item $F^{S} {=} \{(p, (t',t)) {\in} P^{S}{\times} T^{S} \mid (p,t') {\in} F^\sequence \vee (p,t) {\in} F\} \cup \{((t',t),p) {\in} T^S {\times} P^S \mid \allowbreak (t',p) \allowbreak {\in} F^\sequence \allowbreak \vee (t,p) {\in} F\}$

\item $\marking_i^S {=} [p_i^\sequence,p_i]$ and $\marking_f^S {=} [p_o^\sequence,p_o]$

\item $\lambda^S: T^S {\rightarrow} (\univAct {\cup} \{\tau\} {\cup} \{\gg\}) \times (\univAct {\cup} \{\tau\} {\cup} \{\gg\})$ (assuming $\gg {\notin} \univAct {\cup} \{\tau\}$) s.t.:
\begin{itemize}
\item $\lambda^{S}(t',\gg ) {=} (\lambda^\sequence(t'),\gg ) \text{ for } t' {\in} T^\sequence$
\item $\lambda^{S}(\gg ,t) {=} (\gg ,\lambda(t)) \text{ for } t {\in} T$
\item $\lambda^{S}(t',t) {=} \lambda^{S}(\lambda^\sequence(t),\lambda(t)) \text{ for } t' {\in} T^\sequence, t {\in} T$
\end{itemize}
\end{itemize}
\end{definition}

Next, we briefly introduce the shortest path algorithm $A^{\star}$ since our proposed algorithm is based on it and it is widely used for alignment computation~\cite{adriansyah_2014_phd_aligning,DBLP:books/sp/CarmonaDSW18}. 

\subsubsection{$A^{\star}$ algorithm}
The $A^{\star}$ algorithm~\cite{DBLP:journals/tssc/HartNR68} is an informed search algorithm that computes a shortest path. 
It efficiently traverses a search-space by exploiting, for a given state, the \emph{estimated remaining distance}, referred to as the heuristic/$h$-value, to the closest goal state.
The algorithm maintains a set of states of the search-space in its so-called open-set $O$.
For each state in $O$, a path from the initial state to such a state is known and hence, the distance to reach that state, referred to as the $g$ value, is known.
A state from $O$ with minimal $f$-value, i.e., $f{=}g{+}h$, is selected for further analysis until a goal state is reached.
The selected state itself is moved into the closed set $C$, which contains fully investigated states for which a shortest path to those states is known. 
Furthermore, all successor states of the selected state are added to the open set $O$.
Note that the used heuristic must be \emph{admissible}~\cite{DBLP:journals/tssc/HartNR68}. 
If the used heuristic also satisfies \emph{consistency}~\cite{DBLP:journals/tssc/HartNR68}, states need not be reopened. 

\section{Incremental Prefix-Alignment Computation}
\label{sec:algo}
In this section, we present an exact algorithm to incrementally compute optimal prefix-alignments on an event stream. 
First, we present an overview of the proposed approach followed by a detailed description of the main algorithm.

\subsection{Overview}

The core idea of the proposed algorithm is to exploit previously calculated results, i.e., explored parts of the state-space of an SPN.
For each process instance, we maintain an SPN, which is extended as new events are observed.
After extending the SPN, we ``continue'' the search for an optimal prefix-alignment. 

\begin{figure}[tb]
    \centering
    \includegraphics[width=\textwidth,trim=0 10.4cm 3.4cm 0,clip]{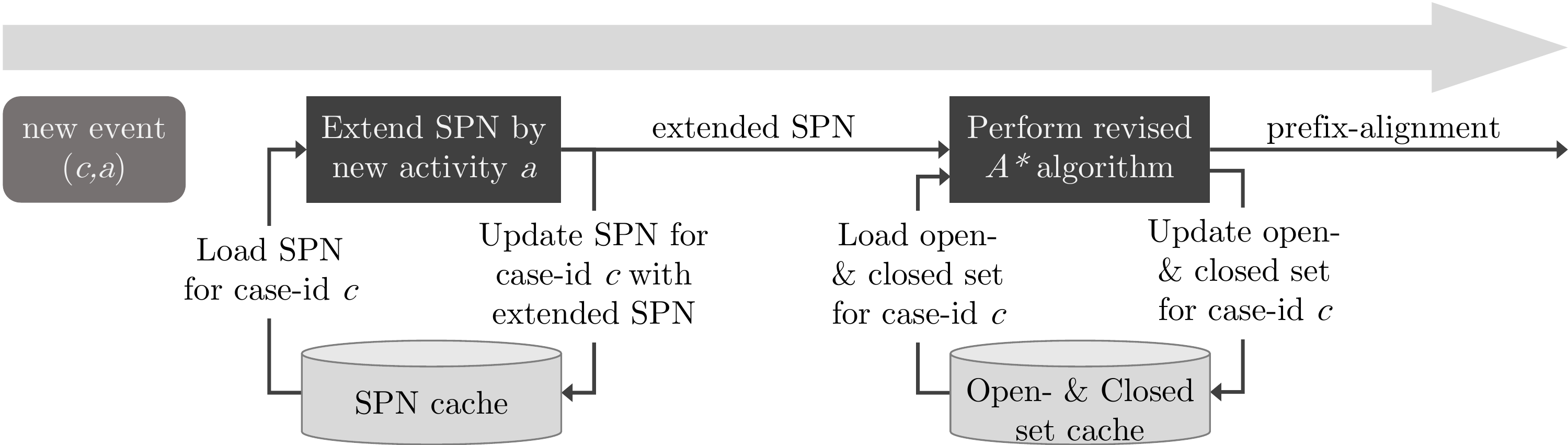}
    \caption{Overview of the proposed incremental prefix alignment approach}
    \label{fig:overview_algo}
\end{figure}

In Fig.~\ref{fig:overview_algo} we visualize a conceptual overview of our approach.
We observe a new event $(\caseId,\activity)$ on the event stream.
We check our SPN cache and if we previously built an SPN for case $\caseId$, we fetch it from the cache.
We then extend the SPN by means of adding activity $a$ to the trace net part.
Starting from intermediate results of the previous search, i.e., open \& closed set used in the $A^{\star}$ algorithm, we find a new, optimal prefix-alignment for case $c$.

\begin{algorithm}
	\footnotesize
	\caption{Incremental Prefix-Alignment Computation}
	\label{alg:inc_prefix}
	\SetKwInOut{Input}{input}
	\Input{${N{=}(\places,\transitions,\pnArcs,[\place_i],[\place_o],\labelFunc)},  
	\stream{\in}(\univCase{\times}\univAct)^*$}
	\Begin{
		\nl \ForAll{$c \in \univCase$}{
    		$\dataStructure_\sequence(c)\gets\langle\rangle$, 
    		$\dataStructure_C(c)\gets\emptyset$\label{alg:line:initialization}\tcp*[r]{initialize cache}
    		
		}
		\nl $i \gets 1$\;
		\nl \While{\texttt{true}}{ \label{alg:line:while_true}
			\nl $\event \gets \stream(i)$\tcp*[r]{get $i$-th event of event stream}
			\nl $\caseId \gets \proj_1(\event)$\tcp*[r]{extract case-id from current event}
			\nl $\activity \gets \proj_2(\event)$\tcp*[r]{extract activity label from current event}

			
			\nl $\dataStructure_{\sequence}(c) \gets \dataStructure_{\sequence}(c) {\cdot} \langle a \rangle$\tcp*[r]{extend trace for case $c$}
			

			\nl let $N^S{=}(P^S,T^S,F^S,\marking_i^S,\marking_f^S,\lambda^S)$ from $N$ and $\dataStructure_{\sigma}(c)$\tcp*[r]{construct/extend synchronous product net}
			
			\nl let $h: \reachableMarkings(N^S,\marking_i^S){\to}\reals_{\geq 0}$\tcp*[r]{define heuristic function}
			
			\nl let $d: T^S{\to}\reals_{\geq 0}$\tcp*[r]{define standard cost function}
			
			\nl \If(\tcp*[f]{initialization for first run regarding case $c$}){$|\dataStructure_{\sigma}(c)|{=}1$}{
			    
			    \nl $\dataStructure_O(c)\gets\{M_i^S\}$\tcp*[r]{initialize open set }
			    
			    \nl $\dataStructure_g(c)\gets M_i^S {\mapsto} 0$\tcp*[r]{initialize cost-so-far function}
			    
			    \nl $\dataStructure_p(c)\gets M_i^S {\mapsto} (\textit{null},\textit{null})$\tcp*[r]{initialize predecessor function}
			    
			}
			
			\nl $\dataStructure_{\prefixAlignment}(\caseId),\dataStructure_O(\caseId),\dataStructure_C(\caseId),\dataStructure_g(\caseId),\dataStructure_p(\caseId) \gets A^{\star}_{\text{inc}}(N^S,\dataStructure_{\sequence}(\caseId), \dataStructure_O(\caseId), \dataStructure_C(\caseId),$
			$\dataStructure_g(\caseId),\dataStructure_p(\caseId),h,d)$\tcp*[r]{execute/continue shortest path search}
			
				
			\nl $i \gets i + 1$\;
		}	
	}			
\end{algorithm}

In Alg.~\ref{alg:inc_prefix} we depict the overall algorithm. 
As input we assume a reference process model, i.e., a WF-net $N$, and an event stream $S$. 
The algorithm processes every event on the stream $S$ in the order in which they occur.
First, we extract the case id and the activity label from the current event. 
Next we either construct the SPN if it is the first activity for the given case or we extend the previously constructed SPN. 
For the SPN's state space we then define a heuristic function $h$ and the standard cost function $d$.
If we process the first activity for a given case, we initialize the open set $O$ with the SPN's initial marking.
Afterwards, we calculate a prefix alignment by calling a modified $A^*$ algorithm, i.e., $A^{\star}_{\text{inc}}$.
We obtain an optimal prefix-alignment $\prefixAlignment$, open set $O$, closed set $C$, cost-so-far function $g$ and the predecessor function $p$.
The function $g$ assigns to already discovered states the currently known cheapest costs and function $p$ assigns the corresponding predecessor state to reach those.
We cache the results to reuse them when computing the next prefix-alignment upon receiving a new event for the given case.
Afterwards, we process the next event.

Note that, the approach theoretically requires infinite memory since it stores all intermediate results for all occurring cases because in general, we do not know when a case is completed in an online dimension.
However, this is a general research challenge in process mining on streaming data, which is not addressed in this paper.

The following sections are structured according to the overview shown in Fig.~\ref{fig:overview_algo}.
First, we explain the SPN extension.
Subsequently, we present a revised $A^*$ algorithm  to incrementally compute optimal prefix-alignments, i.e., $ A^{\star}_{\text{inc}}$.
Moreover, we present a heuristic function for the prefix-alignment computation.

\subsection{Extending SPNs}

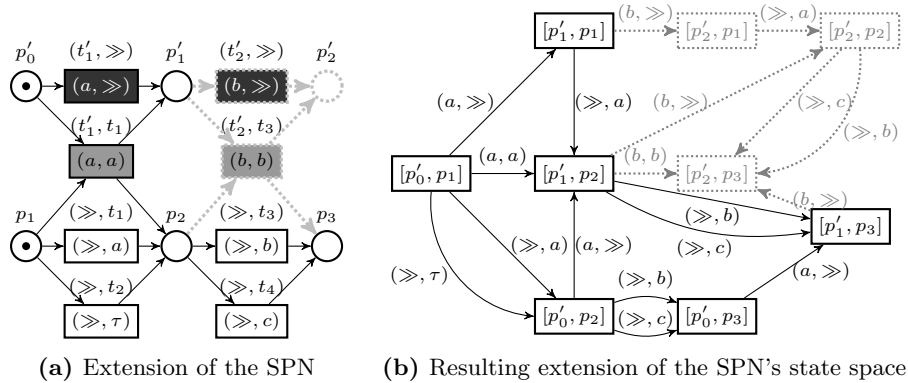
\begin{figure}[tb]
    \centering
    \begin{subfigure}[b]{0.39\textwidth}
        \centering
        \begin{tikzpicture}[node distance=1cm,>=stealth',bend angle=45,auto]
        \scriptsize
    	  \tikzstyle{place}=[circle,thick,draw=black,minimum size=4mm]
    	  \tikzstyle{place_new}=[circle,very thick,draw=lightgray,minimum size=4mm,densely dotted]
    
    	  \tikzstyle{transition}=[thick,draw=black,minimum size=4mm]
    	  \tikzstyle{transition_new}=[very thick,draw=lightgray,minimum size=4mm,densely dotted]
    	  \tikzstyle{silent}=[rectangle,very thick,draw=black,fill=black,minimum size=4mm, text=white]
    	
    	    \node [place,tokens=1,label={$p_0'$}] (p0_) {};
    		\node [transition] (t1_) [right of = p0_, label={$(t_1',\gg)$}, fill=logMove]  {$\color{white}(a,\gg)$};  
    		\node [place,tokens=0,label={$p_1'$}, right of = t1_] (p1_) {};
    		\node [transition_new] (t2_) [right of = p1_, label={$(t_2',\gg)$}, fill=logMove]  {$\color{white}(b,\gg)$};  
    		\node [place_new,tokens=0,label={$p_2'$}, right of = t2_] (p2_) {};
    		\draw [->,] (p0_) to (t1_); 
    		\draw [->,] (t1_) to (p1_); 
    		\draw [->,lightgray,densely dotted,very thick] (p1_) to (t2_); 
    		\draw [-> ,lightgray,densely dotted,very thick] (t2_) to (p2_); 
    		\node [transition,label={$(t_1',t_1)$}, below of = t1_,fill=synchronousMove] (t_aa) {$(a,a)$};
    		\draw [->,] (p0_) to (t_aa); 
    		\draw [->,] (t_aa) to (p1_); 
    
    		\node [transition_new,label={$(t_2',t_3)$}, below of = t2_,fill=synchronousMove] (t_bb) {$(b,b)$};
    		\draw [->,lightgray,densely dotted,very thick] (p1_) to (t_bb); 
    		\draw [->,lightgray,densely dotted,very thick] (t_bb) to (p2_); 
    		
    		
    		\node [place,label={$p_1$}, below = 1.7cm of  p0_,tokens=1] (p1) {};
    		\node [transition] (A) [right of = p1, label={$( \gg,t_1)$}]  {$(\gg,a)$};    
    	    \node [place] (p2) [right of = A,label={$p_2$}] {};
    		\node [transition] (B) [right of=p2,label={$( \gg,t_3)$}]  {$(\gg,b)$};
    		\node [transition] (C) [below of = B,label={$(\gg,t_4)$}]  {$(\gg,c)$};
    		\node [transition] (tau) [below of = A,label={$(\gg,t_2)$}]  {$(\gg,\tau)$};
    
    	    \node [place] (p3) [right of=B,label={$p_3$}] {};
    		\draw [->] (p1) to (A); 
    		\draw [->] (A) to (p2); 
    		\draw [->] (p2) to (B); 
    		\draw [->] (p2) to (C); 
    		\draw [->] (B) to (p3); 
    		\draw [->] (C) to (p3); 
    		
    		\draw [->] (p1) to (t_aa);
    		\draw [->] (t_aa) to (p2);
    		
    		\draw [->,lightgray,densely dotted,very thick] (p2) to (t_bb);
    		\draw [->,lightgray,densely dotted,very thick] (t_bb) to (p3);
    		
    		\draw [->] (p1) to (tau); 
    		\draw [->] (tau) to (p2); 
    		
	    \end{tikzpicture}
        \caption{Extension of the SPN}
        \label{fig:sync_net_ext_spn}
    \end{subfigure}
    \hfill
    \begin{subfigure}[b]{0.59\textwidth}
        \centering
        \begin{tikzpicture}[node distance=1.9cm,>=stealth',bend angle=19,auto]
    	\scriptsize
        \tikzstyle{gray}=[rectangle,draw=black,thick,minimum size=4mm]
      
        \tikzstyle{every label}=[blue]
    
        \node [gray,tokens=0] (1) {$[p_0',p_1]$};
        \node [gray,tokens=0] (2)[right of = 1] {$[p_1',p_2]$};
        \node [gray,tokens=0] (3)[above of= 2] {$[p_1',p_1]$};
    		
        \node [gray,tokens=0, color=gray,densely dotted] (31)[right of= 3] {$[p_2',p_1]$};
    
        \draw [->,color=gray,densely dotted,thick] (3) to node  {$(b,\gg)$} (31);
        \draw [->] (3) to node [right = -0.1cm,text=black] {$(\gg,a)$} (2);
        
        \node [gray,tokens=0] (4)[below of= 2] {$[p_0',p_2]$};
        
        \node [gray,tokens=0] (444)[right of= 4] {$[p_0',p_3]$};
    	\draw [->,bend left] (4) to node  {$(\gg,b)$} (444); 
    	\draw [->,bend right] (4) to node  {$(\gg,c)$} (444); 
    
        \node [gray,tokens=0] (4444)[above right=1cm of 444] {$[p_1',p_3]$};
    	\draw [->] (444) to node [right] {$(a,\gg)$} (4444);

        \node [gray,tokens=0,color=gray,densely dotted] (5)[right of =2] {$[p_2',p_3]$};
       	\draw [->,color=gray,densely dotted,thick] (4444) to node [right] {$(b,\gg)$} (5); 
    
        \node [gray,tokens=0,color=gray,densely dotted] (6)[right of = 31] {$[p_2',p_2]$};
    	
    	\draw [->] (1) to node [text=black] {$(a,a)$} (2); 
    	\draw [->] (1) to node [left,text=black] {$(a,\gg)$} (3);
    	\draw [->] (1) to node [right,text=black] {$(\gg,a)$} (4);  
    	\draw [->, bend right= 45] (1) to node [left,text=black] {$(\gg,\tau)$} (4);

    	\draw [->,color=gray,densely dotted,thick] (2) to node [] {$(b,b)$} (5);  
    	\draw [->,color=gray,densely dotted,thick] (2) to node [left] {$(b,\gg)$} (6);  
    
        \draw [->, bend left=45,color=gray,densely dotted,thick] (6) to node [right] {$(\gg,b)$} (5);
        \draw [->,color=gray,densely dotted,thick] (6) to node [right] {$(\gg,c)$} (5);
        \draw [->,color=gray,densely dotted,thick] (31) to node {$(\gg,a)$} (6); 
        \draw [->] (4) to node [right=-.1cm,text=black] {$(a,\gg)$} (2); 
        
     	\draw [->] (2) to node [below] {$(\gg,b)$} (4444); 
    	\draw [->,bend right] (2) to node [below] {$(\gg,c)$} (4444);

        \end{tikzpicture}
        \caption{Resulting extension of the SPN's state space}
        \label{fig:sync_net_ext_state_space}
    \end{subfigure}
    \caption{Incremental extension of the SPN for the process model $N_1$ and a trace that was extended by a new activity $b$, i.e, $\langle a\rangle {\cdot} \textcolor{gray}{\langle b \rangle}$}
    \label{fig:sync_net_ext}
\end{figure}

Reconsider WF-net $N_1$ (Fig.~\ref{fig:example_petri_net}) and assume that the first activity we observe is $a$.
The corresponding SPN is visualized by means of the solid elements in Fig.~\ref{fig:sync_net_ext_spn} and the state space in Fig.~\ref{fig:sync_net_ext_state_space}.
Any state in the state-space containing a token in $p'_1$ is a suitable goal state of the $A^{\star}$ algorithm for an optimal prefix-alignment.

Next, for the same process instance, we observe an event describing activity $b$.
The SPN for the process instance now describing trace $\langle a,b\rangle$ as well as its corresponding state-space is expanded.
The expansion is visualized in Fig.~\ref{fig:sync_net_ext} by means of dashed elements.
In this case, any state that contains a token in $\place'_2$ corresponds to a suitable goal state of the optimal prefix-alignment search.

\subsection{Incrementally Performing Regular $A^{\star}$}
\label{sec:revised_a_star}
Here, we present the main algorithm to compute prefix-alignments on the basis of previously executed instances of the $A^{\star}$ algorithm. 

\begin{algorithm}[tb!]
	\footnotesize
	\caption{$ A^{\star}_{\text{inc}}$ (modified $A^*$ algorithm that computes prefix-alignments from pre-filled open and closed sets)}
	\label{alg:a_star_inc}
	\SetKwInOut{Input}{input}
	\Input{${N^S{=}(P^S,T^S,F^S,\marking_i^S,\marking_f^S,\lambda^S)},\allowbreak  	
	O,C{\subseteq} \reachableMarkings(N^S,\marking_i^S),\allowbreak
	g {\colon} \reachableMarkings(N^S,\marking_i^S) {\to} \reals_{\geq0},\allowbreak 
	p {\colon} \reachableMarkings(N^S,\marking_i^S) {\to} T^S {\times} \reachableMarkings(N^S,\marking_i^S),\allowbreak
	h {\colon} \reachableMarkings(N^S,\marking_i^S) {\to} \reals_{\geq0},\allowbreak
	d {\colon} T^S {\to} \reals_{\geq0}
	$}
	\Begin{
	\nl let $p_{|\sequence|}$ be the last place of the trace net part of $N^S$\;
	\nl\ForAll(\tcp*[f]{initialize undiscovered states}){$m {\in} \reachableMarkings(N^S,\marking_i^S) {\setminus} O {\cup} C$ \label{alg:line:initialize_undiscovered_states}}{
	    \nl $g(m)\gets\infty$\;
	    \nl $f(m)\gets\infty$\;
	}
	\nl \ForAll{$m {\in} O$}{ 
	    \nl $f(m)=g(m)+h(m)$\tcp*[r]{recalculate heuristic and update $f$-values} \label{alg:line:recalculate_heuristic}
	}
	\nl \While{$O {\neq} \emptyset$}{\label{alg:line:while}
	    \nl $m \gets \underset{m {\in} O}{\arg\min}\medspace f(m)$\label{alg:line_pop_from_o}\tcp*[r]{pop a state with minimal $f$-value from $O$}
	    \nl \If{$p_{|\sequence|} {\in_+} m$}{\label{alg:line:check_goal}
	        \nl $\prefixAlignment \gets $ prefix-alignment that corresponds to the sequence of transitions $(t_1,...,t_n)$ where $t_n{=}\pi_1(p(m)), t_{n-1}{=} \pi_1(\pi_2(p(m))),$ etc. until there is a marking that has no predecessor, i.e., $M_i^S$\;
	        \nl \textbf{return} $\prefixAlignment,O,C,g,p$\;
	    }
	    \nl $C\gets C {\cup} \{m\}$\;
	    \nl $O \gets O {\setminus} \{m\}$\;
	    \nl \ForAll(\tcp*[f]{investigate successor states}){$t {\in} T^S$ s.t. $(N^S,m)[t\rangle(N^S,m')$}{
	        \nl \If{$m' {\notin} C$}{
	            \nl $O\gets O {\cup} \{m'\}$\;
	            \nl \If(\tcp*[f]{a cheaper path to $m'$ was found}){$g(m)+d(t)<g(m')$}{
	                \nl $g(m')\gets g(m)+d(t)$\tcp*[r]{update costs to reach $m'$}
	                \nl $f(m')\gets g(m')+h(m')$\tcp*[r]{update $f$-value of $m'$}
	                \nl $p(m')\gets (t,m)$\tcp*[r]{update predecessor of $m'$}
	            }
	        }
	    }
	}
	}	
\end{algorithm}

The main idea of our approach is to continue the search on an extended search space. 
Upon receiving a new event $(\caseId,\activity)$, we apply the regular $A^{\star}$ algorithm using the cached open- and closed-set for case identifier $\caseId$ on the corresponding extended SPN.
Hence, we incrementally solve shortest path problems on finite, fixed state-spaces by using the regular $A^{\star}$ algorithm with pre-filled open and closed sets from the previous search.
Note that the start state remains the same and only the goal states differ in each incremental step.

In Alg.~\ref{alg:a_star_inc}, we present an algorithmic description of the $A^{\star}$ approach.
The algorithm assumes as input an SPN, the open- and closed-set of the previously executed instance of the $A^{\star}$ algorithm, i.e., for the process instance at hand, a cost-so-far function $g$, a predecessor function $p$, a heuristic function $h$, and a cost function $d$ (standard cost function).
First, we initialize all states that have not been discovered yet (line~\ref{alg:line:initialize_undiscovered_states}).
Since the SPN is extended and the goal states are different with respect to the previous run of the algorithm for the same process instance, all $h$-values are outdated.
Hence, we recalculate the heuristic values and update the $f$-values for all states in the open set (line~\ref{alg:line:recalculate_heuristic}) because we are now looking for a shortest path to a state that has a token in the newly added place in the trace net part of the SPN.
Hence, the new goal states were not present in the previous search problem.
Note that the $g$ values are not affected by the SPN extension.
Thereafter, we pick a state from the open set with smallest $f$-value (line~\ref{alg:line:while}).
First, we check if the state is a goal state, i.e., whether it contains a token in the last place of the trace net part (line \ref{alg:line:check_goal}). 
If so, we reconstruct the sequence of transitions that leads to the state, and thus, we obtain a prefix-alignment (using predecessor function $p$).
Otherwise, we move the current state from the open- to the closed set and examine all its successor states. 
If a successor state is already in the closed set, we ignore it. 
Otherwise, we add it to the open set and update the $f$-value and the predecessor state stored in $p$ if a cheaper path was found.

\subsubsection{Heuristic for Prefix-Alignment Computation}
Since the $A^\star$ algorithm uses a heuristic function to efficiently traverse the search space, we present a heuristic for prefix-alignment computation based on an existing heuristic~\cite{adriansyah_2014_phd_aligning} used for conventional alignment computation. 
Both heuristics can be formulated as an Integer Linear Program (ILP).
Note that both heuristics can also be defined as a Linear Program (LP) which leads to faster calculation but less accurate heuristic values. 

Let $N^S {=} (P^S,T^S,F^S,\marking_i^S,\marking_f^S,\lambda^S)$ be an SPN of a WF-net $\petriNet {=} (\places,\transitions,\pnArcs,[\place_i],\allowbreak [\place_o],\allowbreak\labelFunc)$ and a trace $\sequence$ with corresponding trace net $N^{\sequence} {=} (P^{\sequence},T^{\sequence},F^{\sequence},[p_i^{\sequence}],[p_o^{\sequence}],\allowbreak\lambda^{\sequence})$. 
Let $c\colon T^S {\to} \reals_{\geq 0}$ be a cost function that assigns each (prefix-)alignment move, i.e., transition in the SPN, costs. 
We define a revised heuristic function for prefix-alignment computation as an ILP:
\begin{itemize}
\item \emph{Variables}: $X {=} \{x_t \mid t {\in} T^{S}\}$ and  $\forall x_t {\in} X: x_t {\in} \naturalsZero$
\item \emph{Objective function}: $ min \sum_{t {\in} T^{S}} x_t {\cdot} c(t) $
\item \emph{Constraints}:
\begin{itemize}
    \item Trace net part:
    $\marking_f^{S}(p){=}\sum_{t \in \bullet p} x_t - \sum_{t \in p\bullet} x_t \ \ \forall p {\in} P^{S}: p {\in} P^{\sigma}$ 
    \item Process model part:
    $0 {\leq} \sum_{t \in \bullet p} x_t - \sum_{t \in p\bullet} x_t \ \ \forall p {\in} P^{S}: p {\in} P$ 
\end{itemize}

\end{itemize}

The revised heuristic presented is a relaxed version of the existing heuristic used for conventional alignment computation.
Admissibility and consistency can be proven in a similar way as for the existing heuristic. We refer to~\cite{adriansyah_2014_phd_aligning,DBLP:books/sp/CarmonaDSW18}.

\subsubsection{Reducing Heuristic Recalculations}
In this section, we describe an approach to reduce the number of heuristic calculations. 
Reconsider line \ref{alg:line:recalculate_heuristic} in Algorithm \ref{alg:a_star_inc}.
Before we continue the search on an extended search space, we recalculate the heuristic for all states in the open set.
This is needed because the goal states differ in each incremental execution.
However, these recalculations are computational expensive.
Instead of recalculating the heuristic in advance (Algorithm \ref{alg:a_star_inc}, line \ref{alg:line:recalculate_heuristic}), we mark all states in the open set that have an outdated heuristic value.
Whenever we pop a state from the open set with an outdated heuristic value (line \ref{alg:line_pop_from_o}), we update its $h$-value, put it back in the open set and skip the remainder of the while body (from line \ref{alg:line:check_goal}).
Thereby, we do not have to recalculate the heuristic value for all states in the open set.
This approach is permissible because the goal states are ``further away'' in each iteration and hence, $h$-values can only grow.

\section{Correctness}
\label{sec:correctness}
In this section, we prove the correctness of the approach. 
We show that states in the closed set do not get new successors upon extending the SPN.
Furthermore, we show that newly added states never connect to ``older'' states. 
Finally, we show that the open set always contains a state which is part of an optimal prefix-alignment of the extended trace.


\begin{lemma}[State-space growth is limited to frontier]\label{lem:state_space_frontier}
Let $\sigma^{i-1}{=}\langle a_1,\dots,\allowbreak a_{i-1}\rangle$,\allowbreak $\sigma^i{=}\sigma^{i-1} \cdot \langle a_i\rangle$,\allowbreak and $\sigma^{i+1} {=} \sigma^i {\cdot} \langle a_{i+1}\rangle$. 
For a WF-net, $\petriNet$ let $N_{i-1}^S{=}(P_{i-1}^S,\allowbreak T_{i-1}^S,\allowbreak F_{i-1}^S,\allowbreak \marking_{i_{i-1}}^S,\allowbreak \marking_{f_{i-1}}^S,\allowbreak \lambda_{i-1}^S)$ be the SPN of $N$ and $\sigma^{i-1}$, $N_{i}^S$ and $N_{i+1}^S$ analogously. 

$$\forall \marking {\in} \mathcal{B}(P^S_{i-1}) \forall t {\in} T^S_{i+1}  \left( (N^S_{i+1},\marking)[t\rangle \Rightarrow t {\in} T^S_{i} \right)$$

\begin{proof}[By construction of the SPN]
Observe that $P_{i-1}^S{\subset}P_{i}^S{\subset}P_{i+1}^S$ and $T_{i-1}^S {\subset}T_{i}^S\allowbreak{\subset}T_{i+1}^S$.
Let $p_{|\sigma^i|}{\in}P^S_{i+1}$ be the $i$-th place of the trace net part (note that $p_{|\sigma^i|}{\notin}P_{i-1}^S$) and let $t_{i+1}{\in} T^S_{i+1} {\setminus} T^S_{i}$.
By construction of the SPN, we know that $p_{|\sigma^i|} {\in} {\bullet} t_{i+1}$ and $\forall j {\in} \{1,\dots,i-1\}: p_{|\sigma^j|} {\notin} {\bullet} t_{i+1}$.\qed
\end{proof}
\end{lemma}

Observe that, when searching for an alignment for $\sequence^i$, Alg.~\ref{alg:a_star_inc} returns whenever place $\place_{\sequence^{i}}$ is marked.
Moreover, the corresponding marking remains in $O$.
Hence, each state in $C$ is ``older'', i.e., already part of $\places^S_{i-1}$.
Thus, Lemma~\ref{lem:state_space_frontier} proves that states in the closed set $C$ do not get new successors upon extending the SPN.

\begin{lemma}[New states do not connect to old states] \label{lem:way_back}
Let $\sigma^{i} {=} \langle a_1,\dots,a_{i}\rangle$ and $\sigma^{i+1} {=} \sigma^{i} {\cdot} \langle a_{i+1}\rangle$. 
For a given WF-net $N$, let $N^S_{i} {=} (P^S_i,T^S_i,F^S_i,\marking^S_{i_i},\marking^S_{f_i},\lambda^S_i )$ (analogously $N^S_{i+1}$) be the SPN of $N$ and $\sequence^{i}$.

$$ \forall \marking {\in} \univMSet(P^S_{i+1}) {\setminus} \univMSet(P^S_i)   \forall \marking' {\in} \univMSet(P^S_i)  \left( \nexists t{\in} T^S_{i+1} \left( (N^S_{i+1},\marking) [t\rangle (N^S_{i+1},\marking') \right)\right)$$

\begin{proof}[By construction of the SPN]
Let $t_{i+1} \in T_{i+1}^S{\setminus} T_{i}^S$. Let $p_{|\sigma^{i+1}|}{\in} P^S_{i+1}$ be the $(i+1)$-th place  (the last place) of the trace net part. 
We know that $p_{|\sigma^{i+1}|} {\in} t_{i+1}{\bullet}$ and $p_{|\sigma^{j}|} {\notin} t_{i+1}{\bullet}  \ \forall j {\in} \{1,\dots,i\}$. For all other $t {\in} T_i^S$ we know that $\nexists  \marking{\in} \mathcal{B}(P^S_{i+1}) {\setminus} \mathcal{B}(P^S_{i}) \text{ such that } (N^S,\marking)[t\rangle$.\qed
\end{proof}
\end{lemma}

From Lemma~\ref{lem:state_space_frontier} and \ref{lem:way_back} we know that states in the closed set are not affected by extending the SPN. 
Hence, it is feasible to continue the search from the open set and to not reconsider states which are in the closed set. 

\begin{lemma}[Exists a state in the $O$-set that is on the shortest path]
Let $\sequence^i=\langle a_1,...,a_i \rangle$, $\sequence^{i+1} {=} \sequence^i\cdot \langle a_{i+1}\rangle$, $N^S_i$, $N^S_{i+1}$ the corresponding SPN for a WF-net $N$, $O^i$ and $C^i$ be the open- and closed set after the prefix-alignment computation for $\sigma_i$. Let $\prefixAlignment_{i+1}$ be an optimal prefix-alignment for $\sigma_{i+1}$. 
$$\exists j {\in} \{1,\dots,|\prefixAlignment_{i+1}|\}, \prefixAlignment_{i+1}' {=} (\prefixAlignment_{i+1}(1),\dots,\prefixAlignment_{i+1}(j)) \text{ s.t.}$$
$$ (N^S_{i+1}, \marking^S_{i+1}) {\xrightarrow{\proj^*_2(\prefixAlignment_{i+1}')_{\downarrow_{\transitions}}}} (N^S_{i+1}, \marking_O) \text{ and } \marking_O {\in} O^i$$
\begin{proof}
$\prefixAlignment^{i+1}$  corresponds to a sequence of markings, i.e., $S {=} (\marking_{i+1}^S,\dots,\allowbreak  \marking',\allowbreak\marking'',\allowbreak...,\allowbreak \marking''')$. Let $X^{i+1}{=} \mathcal{B}(P^S_{i+1}) {\setminus} C^i {\cup} O^i$.
It holds that $X^{i+1} {\cap} O^i {=} X^{i+1}{\cap} C^i {=} O^i {\cap}  C^i \allowbreak{=} \emptyset$.
Note that $\marking''' {\in}  X^{i+1}$ because $\marking'''{\notin} \mathcal{B}(P^S_i)$. 
Assume $\forall \marking {\in} S: \marking {\notin} O^i {\Rightarrow} \forall \marking {\in} S: \marking {\in} C^i {\cup} X^{i+1}$. 
Observe that $\marking_i^S {=} \marking_{i+1}^S {\in} C^i$ since initially $\marking_i^S {\in} O^0$and in the very first iteration $\marking_i^S$ is selected for expansion because it is not a goal state, Algorithm \ref{alg:a_star_inc}. 
We know that for any state pair $\marking',\marking''$ it cannot be the case that $\marking' {\in} C^i, \marking'' {\in} X^{i+1}$. 
Since we know that at least $\marking_i^S {\in} C^i$ and $\marking'''{\in}  X_c^{i+1}$ there $\exists \marking',\marking'' {\in} S$ such that $\marking' {\in} C^i, \marking''{\in} O^i$.\qed
\end{proof}
\end{lemma}

Hence, it is clear from Lemma 1-3 that incrementally computing prefix-alignments, continuing the search from the previous open- and closed set, leads to optimal prefix-alignments.

\section{Evaluation}
\label{sec:eval}

We evaluated the algorithm on publicly available real event data from various processes. 
Here, we present the experimental setup and discuss the results.

\subsection{Experimental Setup}
The algorithm introduced in~\cite{vanZelst2017_ijdsa} serves as a comparison algorithm.
We refer to it as Online Conformance Checking (OCC).
Upon receiving an event, OCC partially reverts the previously computed prefix-alignments (using a given maximal window size) and uses the corresponding resulting state of the SPN as start state.
Hence, the algorithm cannot guarantee optimality, i.e., it does not search for a global optimum. 
However, OCC can also be used without partially reverting, i.e., using window size $\infty$.
Hence, it naively starts the computation from scratch without reusing any information, however, optimality is then guaranteed.
We implemented our proposed algorithm, incremental $A^{\star}$ (IAS), as well as OCC in the process mining library \emph{PM4Py}~\cite{ICPM2019_PM4Py_BertiZA}. 
The source code is publicly available\footnote{\url{https://github.com/fit-daniel-schuster/online_process_monitoring_using_incremental_state-space_expansion_an_exact_algorithm}}.
Although, the OCC algorithm was introduced without a heuristic function~\cite{vanZelst2017_ijdsa}, it is important to note that both algorithms, IAS and OCC, use the previously introduced heuristic in the experiments to improve objectivity. 

We use publicly available datasets capturing the execution of real-life processes~\cite{ccc_19,receipt,sepsis,hospital_billing,bpi_ch_19}.  
To mimic an event stream, we iterate over the traces in the event log and emit each preformed activity as an event. 
For instance, given the event log $L{=}[\langle a,b,c \rangle, \langle b,c,d \rangle, \dots]$, we simulate the event stream $\langle (1,a),\allowbreak(1,b),\allowbreak(1,c),\allowbreak(2,b),\allowbreak(2,c),\allowbreak(2,d),\dots \rangle$.
For all datasets except CCC19~\cite{ccc_19} that contains a process model, we discovered reference process models with the Inductive Miner infrequent version (IMf)~\cite{DBLP:conf/bpm/LeemansFA13} using a high threshold. 
This results in process models that do not guarantee full replay fitness.
Moreover, the discovered process models contain choices, parallelism and loops.

\subsection{Results}

\begin{table}[t]
\caption{Results of the conducted experiments for various real-life event logs}
\label{tab:results}
\def\arraystretch{1.25}
\scriptsize
\centering
\setlength{\tabcolsep}{0.005\textwidth}

    \rowcolors{3}{lightgray}{white}
    
    \resizebox{\textwidth}{!}{%
    \begin{tabular}{P{.13\textwidth}?P{.06\textwidth}|p{.06\textwidth}|p{.06\textwidth}|p{.06\textwidth}|p{.06\textwidth}|p{.06\textwidth}|p{.06\textwidth}?p{.06\textwidth}|p{.06\textwidth}|p{.06\textwidth}|p{.06\textwidth}|p{.06\textwidth}|p{.06\textwidth}|p{.06\textwidth}}
    \hline
    \multirow{2}{*}{\textbf{Event log}} & \multicolumn{7}{p{.36\textwidth}?}{\textbf{$\approx$ avg. queued states per trace}} & \multicolumn{7}{p{.36\textwidth}}{\textbf{$\approx$ avg. visited states per trace}} \\ 
                                        & IASR & IAS     & OCC     & OCC-W1     & OCC-W2    & OCC-W5   & OCC-W10    
                                        & IASR & IAS     & OCC     & OCC-W1     & OCC-W2     & OCC-W5  & OCC-W10       \\ \hline
        CCC 19 \cite{ccc_19}    & 774 & 766 & 14614 & 312 & 431 & 885 & 1622
                                & 756 & 751 & 12557 & 212 & 283 & 506 & 932 \\ 
        
        Receipt \cite{receipt}  & 31 & 29 & 65 & 37 & 50 & 82 & 104
                                & 18 & 17 & 26 & 19 & 23 & 33 & 42 \\ 
        
        Sepsis \cite{sepsis}    & 73 & 70 & 532 & 102 & 146 & 285 & 450
                                & 44 & 43 & 232 & 47 & 62 & 103 & 166 \\ 
        
        Hospital \cite{hospital_billing}    & 21& 21 & 42 & 32 & 41 & 65 & 71
                                            & 11 & 11 & 15 & 14 & 17 & 23 & 26 \\ 
        
        BPIC 19 \cite{bpi_ch_19}    & 28 & 28 & 257 & 41 & 57 & 90 & 107    
                                    & 18 & 18 & 154 & 21 & 27 & 40 & 48 \\ \hline

    \end{tabular}
    }%
    \rowcolors{3}{lightgray}{white}
    
    \resizebox{\textwidth}{!}{%
    \begin{tabular}{P{.13\textwidth}?P{.06\textwidth}|p{.06\textwidth}|p{.06\textwidth}|p{.06\textwidth}|p{.06\textwidth}|p{.06\textwidth}|p{.06\textwidth}?p{.06\textwidth}|p{.06\textwidth}|p{.06\textwidth}|p{.06\textwidth}|p{.06\textwidth}|p{.06\textwidth}|p{.06\textwidth}}
    \hline
    \multirow{2}{*}{\textbf{Event log}} & \multicolumn{7}{p{.36\textwidth}?}{\textbf{\# traces with false positives}} & \multicolumn{7}{p{.36\textwidth}}{\textbf{\# variants with false positives}} \\ 
                                        & IASR & IAS     & OCC     & OCC-W1     & OCC-W2    & OCC-W5  & OCC-W10    
                                        & IASR & IAS     & OCC     & OCC-W1     & OCC-W2     & OCC-W5  & OCC-W10     \\ \hline
        CCC 19 \cite{ccc_19}    & 0 & 0 & 0 & 7 & 8 & 1 & 1                           
                                & 0 & 0 & 0 & 7 & 8 & 1 & 1 \\ 
        
        Receipt \cite{receipt}  & 0 & 0 & 0 & 8 & 5 & 3 & 1
                                & 0 & 0 & 0 & 8 & 5 & 3 & 1\\ 
        
        Sepsis \cite{sepsis}    & 0 & 0 & 0 & 59 & 60 & 6 & 1
                                & 0 & 0 & 0 & 58 & 59 & 6 & 1 \\ 
        
        Hospital \cite{hospital_billing}    & 0 & 0 & 0 & 88 & 88 & 69 & 32
                                            & 0 & 0 & 0 & 49 & 49 & 39 & 19 \\ 
        
        BPIC 19 \cite{bpi_ch_19}    & 0 & 0 & 0 & 318 & 259 & 193 & 90    
                                    & 0 & 0 & 0 & 272 & 206 & 145 & 75 \\ \hline
    \end{tabular}
    }%

    \rowcolors{3}{lightgray}{white}

    \resizebox{\textwidth}{!}{%
    \begin{tabular}{P{.13\textwidth}?P{.06\textwidth}|p{.06\textwidth}|p{.06\textwidth}|p{.06\textwidth}|p{.06\textwidth}|p{.06\textwidth}|p{.06\textwidth}?p{.06\textwidth}|p{.06\textwidth}|p{.06\textwidth}|p{.06\textwidth}|p{.06\textwidth}|p{.06\textwidth}|p{.06\textwidth}}
    \hline
    \multirow{2}{*}{\textbf{Event log}} & \multicolumn{7}{p{.36\textwidth}?}{\textbf{$\approx$ avg. computation time (s) per trace}} & \multicolumn{7}{p{.36\textwidth}}{\textbf{$\approx$ avg. number solved LPs (heuristic functions) per trace}} \\ 
                                        & IASR & IAS     & OCC     & OCC-W1     & OCC-W2    & OCC-W5    &OCC-W10
                                        & IASR & IAS     & OCC     & OCC-W1     & OCC-W2     & OCC-W5  & OCC-W10    \\ \hline
        CCC 19 \cite{ccc_19}    & 12.2 & 5.69 & 35.7 & 0.74 & 0.85 & 1.51 & 2.61 
                                & 3345 & 1889 & 8443 & 338 & 393 & 658 & 1066 \\ 
       
        Receipt \cite{receipt}  & 0.12 & 0.04 & 0.05 & 0.04 & 0.04 & 0.07 & 0.09 
                                & 89.2 & 42 & 53 & 40 & 50 & 75 & 91          \\ 
        
        Sepsis \cite{sepsis}    & 0.59 & 0.28 & 0.6 & 0.09 & 0.11 & 0.23 & 0.35 
                                & 518 & 226  & 343 & 104 & 138 & 247 & 356 \\ 
        
        Hospital \cite{hospital_billing}    & 0.05 & 0.03 & 0.03 & 0.02 & 0.03 & 0.04 & 0.05 
                                            & 63 & 30& 35 & 34 & 42 & 61 & 66 \\ 
        
        BPIC 19 \cite{bpi_ch_19}    & 0.4 & 0.19& 0.79 & 0.06 & 0.09 & 0.12 & 0.14    
                                    & 128 & 71& 136 & 44 & 57 & 81 & 91\\ \hline
    \end{tabular}
    }%

\end{table}

In Table~\ref{tab:results}, we present the results.
OCC-W$x$ represents the OCC algorithm with window size $x$,
OCC with an infinite window size.
Moreover, we present the results for the IAS algorithm that does \emph{not} use the approach of reducing heuristic recalculations as presented in Section \ref{sec:revised_a_star}, we call it IASR.
Note that only IAS(R) and OCC guarantee optimality. 
Furthermore, note that a queued state corresponds to a state added to the open set and a visited state corresponds to a state moved into the closed set.
Both measures indicate the search efficiency.

We observe that reducing the number of heuristic re-calculations is valuable and approximately halves the number of solved LPs and hence, reduces the computation time.
As expected, we find no significant difference in the other measured dimensions by comparing IAS and IASR.
We observe that IAS clearly outperforms all OCC variants regarding search efficiency for all used event logs except for CCC19 where OCC variants with small window sizes have a better search efficiency. 
This results illustrate the relevance of IAS compared to OCC and OCC-W$x$ and show the effectiveness of continuing the search on an extended search space by reusing previous results.
Regarding false positives, we observe that OCC-W$x$ variants return non-optimal prefix-alignments for all event logs.
As expected, the number of false positives decreases with increasing window size.
In return, the calculation effort increases with increasing window size.
This highlights the advantage of the IAS' property being parameter-free.
In general, it is difficult to determine a window size because the traces, which have an impact on the ``right'' window size, are not known in an online setting upfront.
Regarding calculation time, we note that the number of solved LPs has a significant influence.
We observe that IAS has often comparable computation time to the OCC-w$x$ versions.
Comparing optimality guaranteeing algorithms (IAS \& OCC), IAS clearly outperforms OCC in all measured dimensions for all logs.
\subsection{Threats to Validity}

In this section, we outline the limitations of the experimental setup.
First, the artificial generation of an event stream by iterating over the traces occurring in the event log is a simplistic approach.
However, this allows us to ignore the general challenge of process mining on streaming data, deciding when a case is complete, since new events can occur at any time on an (infinite) event stream. 
Hence, we do not consider the impact of multiple cases running in parallel.

The majority of used reference process models are discovered with the IMf algorithm.
It should, however, be noted that these discovered models do not contain duplicate labels. 
Finally, we compared the proposed approach against a single reference, the OCC approach. 
To the best of our knowledge, however, there are no other algorithms that compute prefix-alignments on event streams.

\section{Conclusion}
\label{sec:conclusion}
In this paper, we proposed a novel, parameter-free algorithm to efficiently monitor ongoing processes in an online setting by computing a prefix-alignment once a new event occurs. 
We have shown that the calculation of prefix-alignments on an event stream can be ``continued'' from previous results on an extended search space with different goal states, while guaranteeing optimality.
The proposed approach is designed for prefix-alignment computation since it utilizes specific properties of the search space regarding prefix-alignment computation and therefore, generally not transferable to other shortest path problems.
The results show that the proposed algorithm outperforms existing approaches in many dimensions and additionally ensures optimality.


In future work, we plan to implement the proposed approach in real application scenarios and to conduct a case study.
Thereby, we want to focus on limited storage capacities, which requires to decide whether a case is considered to be completed to free storage.
%
%
%
\bibliographystyle{splncs04}
\bibliography{samplepaper}

\end{document}